%% file: samplepaper.tex
\documentclass[runningheads]{llncs}
\usepackage[T1]{fontenc}
\usepackage[notes=true,later=true,llncssubsub]{dtrt}
\usepackage[advantage,adversary,asymptotics,complexity,keys,lambda,landau,logic,sets,mm,oracles,primitives,probability,notions,sets,operators]{cryptocode} 
\usepackage{color, colortbl, tcolorbox}
\usepackage{url, array, cite, multirow, mathtools, framed, newfloat,microtype,algpseudocode,algorithm}
\usepackage[T1]{fontenc}
\usepackage{diagbox, booktabs, appendix, alltt}
\usepackage{listings, xcolor, afterpage, bm, xspace}
\usepackage[inkscapelatex=false]{svg}
\usepackage[shortlabels]{enumitem}
\usepackage[multiple]{footmisc}
\usepackage[cal=boondox]{mathalfa}
\usepackage[capitalise]{cleveref}
\usepackage{stmaryrd}
\usepackage{caption}
\usepackage{tcolorbox}
\usepackage[english]{babel}
\usepackage{comment}
\usepackage{subcaption}
\usepackage[mathscr]{euscript}
\usepackage{tabularx}
\usepackage{tikz}
\usetikzlibrary{arrows.meta, positioning, fit}
\usepackage{centernot}

\usepackage{soul}

\usepackage{graphicx}

\input{macros}

\begin{document}
\title{All Proof of Work But No Proof of Play}
%
%
\author{Hayder Tirmazi}
%
%
\institute{City College of New York \\
\email{hayder.research@gmail.com}}
\maketitle              
%


\section{Introduction}

Speedrunning is a competition that emerged from communities of early video games such as Doom (1993)~\cite{inverseCoinedspeedrunning}. Speedrunners try to finish a game in minimal time~\cite{penguinz0SpeedRunningChallenge}. Provably verifying the authenticity of submitted speedruns is an open problem. Traditionally, best-effort speedrun verification is conducted by on-site human observers, forensic audio analysis~\cite{arstechnicaScourgeCheating}, or a rigorous mathematical analysis of the game mechanics\footnote{For example, the 29-page official report~\cite{dreamCheating} for the Dream Minecraft case~\cite{gamersFraud} rigorously proves that the probability of the investigated speed run \textit{not} being fraudulent is $\leq 1.326 \times 10^{-13}$, with probabilities taken over the random coins used in item bartering and other game mechanics.}. Such methods are tedious, fallible, and, perhaps worst of all, not cryptographic. Motivated by naivety and the Dunning-Kruger effect, we attempt to build a system that cryptographically proves the authenticity of speedruns. This paper describes our attempted solutions and ways to circumvent them. Through a narration of our failures, we attempt to demonstrate the difficulty of authenticating live and interactive human input in untrusted environments, as well as the limits of signature schemes, game integrity, and provable play.\medskip

\noindent\textbf{Running Example.} We will use an open-source video game~\cite{lilypondGithub} we developed as a running example to both demonstrate techniques for generating fraudulent speedruns and our attempts at cryptographically verifying speed run authenticity. Our video game, lilypond, is developed in roughly 7K lines of \texttt{C} code. We have also released a web version of lilypond that can be played by readers directly on their web browser~\cite{lilypondItch}. The game has a playable character called Mano who can move left and right, jump, sprint, and defeat enemies by jumping on top of them, i.e, \textit{squishing} them. Players control Mano's movements using the keyboard. Mano can also collect coins by colliding with them. If Mano collides with an enemy, such as a bug or a ghost, she is relocated back to the start of the level. If Mano falls into water or lava, she is also relocated to the start.\medskip

\noindent\textbf{Notation.} We introduce the notation we will be using throughout the paper. $\adv$ denotes a $\ppt$ adversary. $\sigma_{t} \in \Sigma$ denotes the internal state of the game at time $t$, where $\Sigma$ is the set of all such states. In Lilypond, this is the struct containing Mano's co-ordinators, current position, etc., and a level struct containing similar information for every other object in the game(Lst.~\ref{lst:state}). $I = \NN \times \mathcal{K} = \{(t_{i}, k_{i})\}$ is a set of timestamped keystrokes, with $t_{i}$ being the frame number, and $\mathcal{K}$ being the set of playable keys in the game (up, down, jump, $\cdots$). $I$ models the interactive inputs received by the game. Pixel depth $n_{p}$ is the number of bits used to encode a pixel on a screen. Using $P = \bin^{n_{p}}$, a screenshot $s \in P^{w h}$ models a single rendering of the game on a screen of $w \times h$ pixels (each $p \in P$ is a single pixel on the screen). A screenshotting function $f : \Sigma \mapsto P^{wh}$ maps a game state to a screenshot. A speedrun $S \subset \NN \times P^{wh} = \{(t_{i}, s_{i})\}$ is a set of timestamped screenshots $s_{i}$. Now we can use our notation to formally define a video game.

\begin{definition}
A $\ppt$ algorithm $G$ is a video game if it interactively takes a set of timestamped keystrokes $I$ and outputs a set of states $\mathscr{S} =\{ \sigma_{0}, \cdots, \sigma_n \}$ using a transition function $T: \NN \times \NN \times \mathcal{K} \times \Sigma \mapsto \Sigma$ such that 
\(
\sigma_{t+1} \leftarrow T(t_{s}, t, \sigma_t, k_t)
\). Note that $t_{s}$ is the system time while $t$ is the in-game (logical) time.
\end{definition}

\begin{figure}[t]
    \centering
    \begin{minipage}[b]{0.48\linewidth} 
        \centering
        \includegraphics[width=\linewidth,trim=0 170px 0 0,clip]{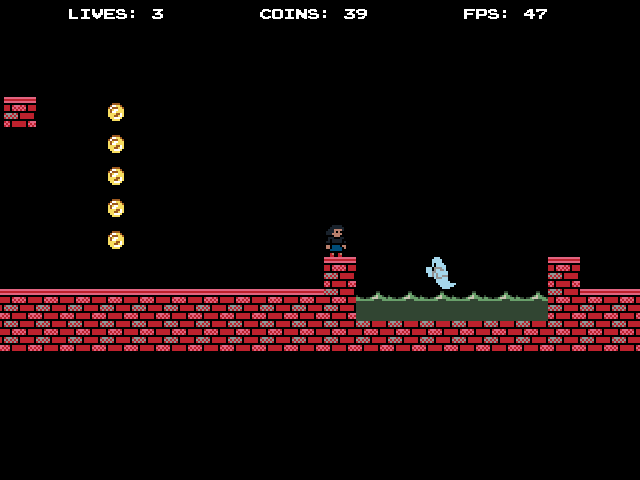}
        \caption{Experimental game lilypond}
        \label{fig:mano1}
    \end{minipage}\hfill 
    \begin{minipage}[b]{0.48\linewidth} 
        \centering
        \includegraphics[width=\linewidth]{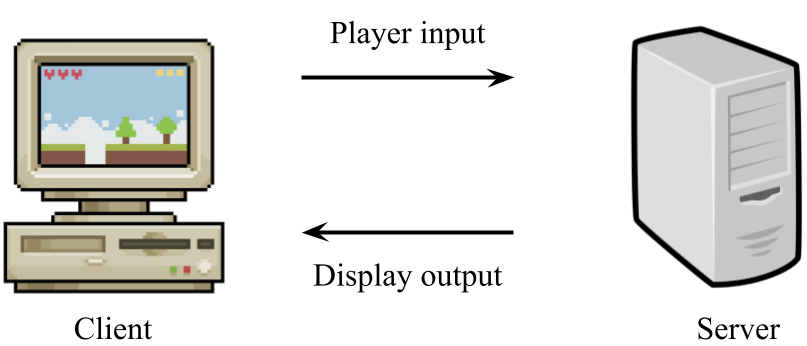}
        \caption{Thin client environment}
        \label{fig:thinclient}
    \end{minipage}
\end{figure}

\noindent\textbf{Fraud Techniques.} We discuss four common methods of speedrun fraud. The most common method is splicing~\cite{cheatingPolygon}. In a lilypond run (Fig.~\ref{fig:mano1}) with a jump, an optimal speedrun involves Mano clearing the jump in one attempt. $\adv$ can make several failed attempts at the jump (falling into the water) and cut that footage out of the final speedrun recording. $\adv$ can also stitch together only the successful attempts for all of Mano's jumps in multiple parts of the game to create a single recording in which Mano appears to clear all jumps successfully in one run.

The second technique is logic modification either in the game binary or at runtime. $\adv$ can modify the game binaries to change the speed at which Mano falls to the ground or Mano's jump height (Lst.~\ref{lst:constants}). $\adv$ can also modify the in-memory game state~\cite{cheatEngine} at runtime (Lst.~\ref{lst:state}) to increase the value of coins collected by Mano or the level Mano is in currently.

The third technique is simulated input. $\adv$ can simulate keyboard input~\cite{arstechnicaScourgeCheating} to send a pre-recorded set $I$ of input values with frame-perfect precision. $I$ may be either algorithmically generated or a replay of a more successful speedrunner. 

The fourth technique is timestamp skew. $\adv$ can slow down in-game timers (Lst.~\ref{lst:timers}) that control Mano's acceleration and other properties, or the system timers that the in-game timers synchronize on. $\adv$ then plays a slowed down version of the game which makes the game significantly easier. $\adv$ speeds up the final speedrun recording so it looks like the game was played at its original pace.

\noindent\textbf{Execution environments.} There are two common execution environments for video games: thin clients~\cite{chang2011understanding, claypool2012thin, lee2015outatime} and thick clients~\cite{lee2015outatime}. Thin clients run the game binary on a cloud server, receive player input over the network, and stream visual output back to the user. The client on the player's device is \textit{thin}, i.e, it only acts as an input collector and video streamer. Nvidia~\cite{geforceNow}, Sony~\cite{playstationportal}, and Amazon~\cite{amazonluna} all offer such cloud gaming subscriptions, allowing a game to be played entirely on their trusted hardware and the visual and audio output streamed to the player's client. A thick client runs the game binary on the player's device. It may use a cloud server for sending state updates in the case of multiplayer games, but does not execute the game directly on the server.

\section{Security Notions}

We first discuss game execution oracles. Next, we introduce two security games in increasing order of adversarial power. Finally, we introduce a security notion for the games. \medskip

\noindent\textbf{Game Execution Oracles.} A thin client oracle $\oracle_{1}$ for a game $G$ and screenshotting function $f$ takes a timestamped keystroke $(t, k)$ as input and outputs a screenshot $s = f(G(t, k))$ . Thus $\adv$ plays the game without having access to either the internal game state, the system timer, or the screenshotting function. A thick client oracle $\oracle_{2}$ for a game $G$ takes as input all the input values for $G$'s transition function $T$, namely, the system time $t_{s}$, the game time $t$, a keystroke $k_{t}$, and a game state $\sigma_{t}$. $\oracle_{2}$ outputs the game state $\sigma_{t + 1} = T(t_{s}, t, \sigma_{t}, k_{t})$. $\adv$ chooses their own screenshotting function for converting the returned game state to a screenshot.\medskip

\noindent\textbf{Security Games.} We now use the game execution oracles to define two security games. For a video game $G$ and $n, \lambda \in \NN$ ($\lambda$ is the security parameter), we define \texttt{ThinClientGame} using a thin client oracle $\oracle_{1}$ as follows.\medskip

\noindent \texttt{ThinClientGame}$(\adv, G, \verifier, t, \lambda)$:
\begin{enumerate}
    \item $\adv$ plays $G$ using $\oracle_{1}$ up to game time $t$ to get screenshot $s_{t}$.
    \item $\adv$ outputs challenge screenshot $s^{*}$
    \item If $\verifier$ accepts $s^{*}$ and $s^{*} \neq s_{t}$, return 1 ($\adv$ wins). Otherwise, return 0.
\end{enumerate}

Similarly, we define \texttt{ThickClientGame} using a think client oracle $\oracle_{2}$ and a screenshotting function $f$ as follows.\medskip

\noindent \texttt{ThickClientGame}$(\adv, G, \verifier, t, \lambda)$:
\begin{enumerate}
    \item $\adv$ plays $G$ using $\oracle_{2}$ up to game time $t$ to get state $\sigma_{t}$.
    \item $\adv$ outputs challenge screenshot $s^{*}$
    \item If $\verifier$ accepts $s^{*}$ and $s^{*} \neq f(\sigma_{t})$, return 1 ($\adv$ wins). Otherwise, return 0.
\end{enumerate}

We know define a security notion for both games.

\begin{definition}
    Let a speedrun construction $C = (G, \verifier, f)$ be a 3-tuple with a game $G$ and corresponding verifier $\verifier$, and screenshotting function $f$. Let $\adv$ be a $\ppt$ adversary. We say $C$ is an $(t, \varepsilon)$-secure construction under a security game \texttt{Game} if 
    \[ \Pr[\texttt{Game}(\adv, G, \verifier, t, \lambda) = 1] \leq \varepsilon \]
    \noindent where $\lambda$ is the security parameter and the probabilities are taken over the random coins of $\adv$ and $G$.
\end{definition}



\noindent\textbf{Thin Client Security}
We show that any speedrun construction $C$ can be transformed (in polynomial time) into a construction $C^{\prime}$ that is secure under \texttt{ThinClientGame} as long as we are allowed to modify least $\frac{n_{p}}{\lambda}$ pixels in every screenshot. Recall that $n_{p}$ is the number of bits per pixel and $\lambda$ is the security parameter. 

\begin{theorem}
    Let $C = (G, V, f)$ be a speedrun construction. Assuming one-way functions, there exists a modified construction $C^{\prime}$ that is $(t, \negl)$-secure for any $t \in \NN$ that modifies at most $\frac{n_{p}}{\lambda}$ pixels in each screenshot $s_{i}$ of the speedrun. 
\end{theorem}
\noindent\textit{Proof Sketch.} Encode signatures with timestamp, input, and game state into pixels of the screenshot, treating them as authenticated logs of play. $\verifier$ checks these signatures in the screenshot.

\begin{proof}
    We modify $G$'s transition function $T$ to create a modified function $T^{\prime}$. Let $\sigma_{t + 1} = T(t_{s}, t, \sigma_{t}, k_{t})$ and $y = t_{s} \| t \| \sigma_{t} \| k_{t}$. Instead of returning $\sigma_{t + 1}$ directly, $T^{\prime}$ returns $\sigma^{\prime}_{t + 1} = (\sigma_{t + 1}, \sign_{\sk}(y))$ where $\sign_{\sk}$ is a digital signature scheme that outputs a signature of length $\lambda$ bits using a private key $\sk$ of length $\bigO{\lambda}$ embedded in the game binary. We modify the screenshotting function $f$ to create a new function $f^{\prime}$ that first invokes $f$ and then writes the signature in $\frac{n_{p}}{\lambda}$ pixel indices chosen beforehand. Note that each screenshot has $w \times h \times n_{p}$ bits, so encoding the $\lambda$ bits long signature means modifying $\frac{n_{p}}{\lambda}$ pixels. We modify the verifier $\verifier$ to only accept a screenshot if the values of the chosen pixel indices equal the signature value. Suppose $\adv$ outputs a screenshot $s^{*} \neq s_{t}$ that $\verifier$ accepts. Then $s^{*}$ must contain a valid signature on some $y^{\prime} \neq y$, i.e., a forgery. Since we assume one-way functions exist, the unforgability of $\sign_{\sk}$ implies that the probability of $\adv$ winning the security game is negligible.
    \qed
\end{proof}

Under a thin client, we are naturally immune to logic modification and timestamp skew as $\adv$ does not have access to the game binaries, memory, or system timers. However, thin client games are still vulnerable to splicing and simulated inputs. \texttt{ThinClientGame} and its security notion does succeed in mitigating splicing, but it does not mitigate simulated input. 

\section{Epic Fails \& Open Problems}

We now discuss the failures and shortcomings of our security notions and our fruitless attempts at improving them. We leave some open problems for the community so brighter minds may succeed where we failed.\medskip

\noindent\textbf{Problem 1: security notions covering simulated inputs.} Our security games do not provide any provable guarantees against simulated inputs. We were unable to propose reasonable assumptions for what such a security notion would look like. One notion we thought about was computational indistinguishability. Challenger $\cdv$ generates an honest input $I_{0}$ and asks adversary $\adv$ for a simulated input $I_{1}$. $\cdv$ then flips a bit $b$ and gives $\adv$ the input $I_{b}$. Can we bound $\adv$'s probability of inferring $b$ and, if so, does this provide meaningful security? We leave the formulation of a meaningful security notion that covers simulated inputs as an open problem.\medskip

\noindent\textbf{Problem 2: empirical methods for simulated input detection.} We tried experimentally detecting simulated inputs. We logged timestamped keystrokes received from a human speedrun of lilypond and edited the log to remove suboptimal inputs. For example, the honest speedrun walks Mano in the wrong direction (left key) before turning back. The edited log removes the left keystrokes for the missteps and the extra right keystrokes of the return journey, giving a dishonest speedrun where Mano moves right without the missteps, saving time. We were unable to distinguish this modified speedrun from a second honest speedrun, where the player simply took the correct path the first time.

Discounting trusted hardware, e.g, keys having a biometric fingerprint sensor, we did not find a way to detect a dishonest speedrun compared to repeated honest speedruns with the player organically performing better over time. For other metrics such variations in keypress interarrival time and the entropy of the set of input stream, we found no convincing arguments for why any such metric cannot also be efficiently simulated, especially from training data taken directly from honest speedruns. We leave this as an open problem.
\medskip

\noindent\textbf{Problem 3: provable security, or an impossibility proof, for the thick client setting.} Our attempts at finding a secure construction for \texttt{ThickClientGame} failed. On a thick client, $\adv$ controls everything, including the game binary, memory, system timers, and input hardware. We question whether meaningful security guarantees under such assumptions are even possible.\medskip

\noindent\textbf{Problem 4: new execution environments and security games.} While our security games are based on execution environments deployed today, they seem to be difficult to achieve guarantees for. \texttt{ThinClientGame} assumes adversary $\adv$ controls only the input. We attempted an even stronger model where $\adv$ cannot manipulate input streams, but that is akin to requiring a player to record their speedruns in Fort Knox. \texttt{ThickClientGame}, on the other hand, assumes $\adv$ controls everything. We also tried making assumptions that were slightly stricter but still realistic, such as game binary signing, but with limited success. We leave this as an open problem.\medskip

\noindent While this paper may serve as proof of work, we were unable to provide a proof of play. We did make some friends along the way. Though they were just bots, running simulated inputs that are doomed to stay.

\bibliographystyle{splncs04}
\bibliography{paper}

\appendix

\section{Code Listings}

\begin{lstlisting}[language=C,caption=Game State,label=lst:state]
struct mano {
  double x, y, vx, vy;
  bool air, ladder, sprint, jump;
  unsigned int lives, coins;
  size_t respawn_x, respawn_y;
};

struct sprite {
  double x, y, vx, vy;
  bool removed;
  union sprite_data data;
};

struct level {
  struct sprite_type **passive_sprites;
  struct array *active_sprites;
};
\end{lstlisting}

\begin{lstlisting}[language=C,caption=Game Physics Constants,label=lst:constants]
static const Sint64 GRAVITY = 600;
static const Sint64 WALK = 72;
static const Sint64 SPRINT = 96;
static const Sint64 LADDER = 64;
static const Sint64 FALL_MAX = 128;
static const Sint64 JUMP = 256;
static const Sint64 SHORT_JUMP = 192;
\end{lstlisting}

\begin{lstlisting}[language=C,caption=Game Timers,label=lst:timers]
struct fps_timer {
  Uint64 delay;
  Uint64 time_left;
};

enum {
  FRAME_RATE = 48,
  FRAME_TIME = 1000 / FRAME_RATE,
  MIN_FRAME_RATE = 24,
  MAX_FRAME_TIME = 1000 / MIN_FRAME_RATE
};
\end{lstlisting}

\end{document}

%% file: macros.tex
\DeclareMathAlphabet{\mathcal}{OMS}{cmsy}{m}{n}

\spnewtheorem{game}{Game}{\bfseries}{}
\spnewtheorem{construction}{Construction}{\bfseries}{}

\spnewtheorem{theorem}{Theorem}{\bfseries}{}

\let\olddefinition\definition
\renewcommand{\definition}{\olddefinition\normalfont}

\graphicspath{{content}}

\definecolor{codegreen}{rgb}{0,0.6,0}
\definecolor{codegray}{rgb}{0.5,0.5,0.5}
\definecolor{codepurple}{rgb}{0.58,0,0.82}
\definecolor{backcolour}{rgb}{1,1,1}

\lstdefinestyle{mystyle}{
    backgroundcolor=\color{backcolour},   
    commentstyle=\color{codegreen},
    keywordstyle=\color{codepurple},
    numberstyle=\tiny\color{codegray},
    stringstyle=\color{codepurple},
    basicstyle=\ttfamily\smaller,
    breakatwhitespace=false,         
    breaklines=true,                 
    captionpos=b,                    
    keepspaces=true,                 
    numbers=left,                    
    numbersep=5pt,                  
    showspaces=false,                
    showstringspaces=false,
    showtabs=false,                  
    tabsize=1
}